\documentclass[twocolumn,twoside,slac_two]{revtex4}
\usepackage{graphicx}
\usepackage{fancyhdr}
\usepackage{hyperref}
\hypersetup{
    colorlinks=true,
    urlcolor=magenta,
}
\begin{document}

\title{High-energy neutrino astronomy with KM3NeT-ARCA}

%

\author{R. Coniglione for the KM3NeT collaboration}
\affiliation{LNS-INFN, Via S. Sofia 62, 95125 Catania, Italy}

\begin{abstract}
The KM3NeT/ARCA high energy neutrino telescope is currently under construction in the Mediterranean sea. The detector will consist of
two blocks of instrumented structures and will have a size of the order of a cubic-kilometer. In this
work the status of the detector, the expected performance to galactic and extragalactic neutrino sources, the results from prototypes and the first deployed lines will be briefly reported.
ì
\end{abstract}

\maketitle

\thispagestyle{fancy}

\section{\label{Introduction}INTRODUCTION}

The KM3NeT collaboration has designed and started the construction of two research infrastructures in the Mediterranean Sea 
hosting two neutrino detectors and nodes for Earth and Sea sciences. The two neutrino detectors share the same technology 
but have two different physics goals. KM3NeT/ARCA, to be installed off shore the Sicilian coast at a depth of 3500~m, 
aims at the discovery and subsequent observation of galactic and extragalactic high-energy neutrino sources while KM3NeT/ORCA, 
to be installed off shore the French coast at a depth of 2500~m, aims at the determination of the neutrino mass
hierarchy by measuring the oscillation of atmospheric neutrinos. 

The main detector elements, namely the optical sensors (Digital Optical Modules - DOM) and the detection units (DU), are identical in the two detectors. The DUs are arranged in blocks of 115.
The granularity of the detectors will be different: the KM3NeT/ARCA detector, when completed, will have an instrumented volume of about 
1 km$^{3}$ while KM3NeT/ORCA, will be smaller and more densely instrumented.
In this work the KM3NeT/ARCA detector will be briefly described and the current status and the 
expected performance of the complete detector reported. The KM3NeT/ORCA detector is discussed in~\cite{ORCA}.

A detailed description of the technical design of the two detectors and their performances are reported in the KM3NeT Letter Of Intent~\cite{LoI}.
 
\section{\label{Physics}Physics objectives and expected performances}
The IceCube observation of a high-energy astrophysical neutrino flux is nowadays very well assessed. 
Different analyses, with sophisticated and different methodologies, have proven the existence of an extraterrestrial neutrino flux~\cite{HESE-upgoing,HESE-combined,HESE-UltraHighEnergy}. 
This discovery represents a turning point in the astroparticle physics. 
In fact, neutrinos are a unique probe for the investigation of the hadronic or leptonic origin of the high energy cosmic rays present in the Universe.  

With the available data the origin of these neutrinos has not yet  been
determined and the spectral shape and flavour composition have not been precisely defined. The measurement of the 
diffuse neutrino flux measured by IceCube with a detector, like ARCA, with a different field of view and a better angular resolution can shed light on the origin of these neutrinos. 
Moreover, the large visibility of the galactic plane and the good angular resolution, both for track and cascade events, makes ARCA the ideal instrument to investigate galactic sources and in general all sources located in the southern sky. 

The performance of the full ARCA detector (2 building blocks of 115 DUs each) to diffuse and point-like sources has been extensively studied by means of Monte Carlo simulations and the results are reported in Ref.~\cite{LoI}. 
In the Monte Carlo simulations the detector response to particles crossing the
detector, their interaction with the medium surrounding the detector and subsequent Cherenkov light production,
and the detector response in terms of the PMT data sent to shore have been taken into account. 
NC and CC interactions for all the three neutrino flavors have been simulated. Background events of atmospheric muons with a live time of about three years and atmospheric neutrinos have also been simulated.

The sensitivity of the ARCA detector to the neutrino flux measured by IceCube has
been evaluated using both track-like events reconstructed up to 10$^\circ$ above the horizon and cascade-like
events in the full angular range. The results of this analysis show that a 
significance of 5$\sigma$ can be reached in less than one year.

The most intense high energy gamma ray sources and generic point-like neutrino emitters have been also considered for the estimate of the discovery potential in ARCA. 
In particular, the Galactic sources RXJ1713 and Vela X have been considered. The sources have been simulated as extended homogeneous sources with the same radius observed in the high energy gamma measurements. Neutrino spectra with energy cutoffs of few TeV have been also assumed. Results from simulations show that a significance of about
3$\sigma$ is reached in 4 years for the RXJ1713 and in about 2 years for the VelaX~\cite{LoI}.
Generic point-like neutrino sources with a neutrino energy spectrum proportional to  E$^{-2}$ have been considered and compared with the correspondent IceCube curve. 

Recently, a new track reconstruction with improved angular resolution ($<0.1^\circ$ for E$_\nu>100$ TeV and near the intrinsic $\nu-\mu$ angle for E$_\nu<10$ TeV) and efficiency has been developed inside the collaboration. Moreover, a more sophisticated multivariate analysis has been applied in the analysis of point-like sources to discriminate signal events from background events. In the neutrino telescopes two different background events  with different angular distributions and energy spectra are present: the atmospheric neutrino events and the atmospheric muon events. We have recently applied a Random Forest algorithm to our simulated data, able to distinguish between three different samples: two backgrounds samples and a signal sample. The first very preliminary results show that an improving of about one year in the RXJ1713 discovery at 3$\sigma$ is achieved.

\section{\label{Design}Detector design}
The observation of high energy neutrinos is possible if particles produced in the rare neutrino interactions with the matter surrounding the 
instrumented volume are detected. To reconstruct the tracks of the secondary particles produced, and from that also the neutrino
direction and its energy, the arrival times of the light collected by optical sensors and their geometrical positions are measured. 

The optical sensors are hosted in flexible string-like structures, called Detection Units (DUs), anchored at the sea floor, kept vertical by buoys and 
connected to shore by an electro-optical cable network.
KM3NeT/ARCA consists in two building blocks of 115 DUs each one hosting 18 optical sensors, called Digital Optical Modules (DOM). 
The horizontal distance between the DUs is about 90~m and the vertical distance between adjacent DOMs is 36~m. 

Both the DOM and the DU are technologically innovative elements designed and constructed inside the KM3NeT collaboration.
The DOM consists of a 17-inch diameter pressure resistant glass sphere that contains 31 3-inches photomultipliers with low power high-voltage bases 
and the readout electronics. 19 PMTs are located in the lower hemisphere and are down-looking while the remaining 12 PMTs are located in the upper hemisphere and are up-looking. The DOM contains also three calibration sensors: a LED nano-beacon for time calibration, a compass
and tiltmeter for orientation calibration and an acoustic piezo sensor, glued to the inner surface, for position calibration.

The DU consists in two parallel ropes to which the DOMs are fixed via a titanium collar. For the power and data transmission a vertical electro-optical
cable, a pressure balanced, oil-filled, plastic tube that contains two copper wires and 18 optical fibres, is attached to the ropes. 
At each DOM two power conductors and a single fibre are branched out via the breakout box.

For the deployment, each detection unit is wound around a spherical frame with diameter of about
2.2~m (Launcher of Optical Modules, LOM), which is positioned on the seabed and then unwinded with a rotating upwards movement. At the end of operation
the LOM reaches the sea surface, where it is collected for reuse.

Power and data are transferred to/from the seafloor infrastructure via a main electro-optic cable. The readout of the detectors is based on the ``All-data-to-shore'' concept.

The new technical designs have been validated through different prototypes.
A DOM prototype hosted on the ANTARES instrumented line has been deployed in May 2013 and operated for one year. 
The results, reported in~\cite{PPM-DOM}, have proven the validity of the design and the powerful of the single DOM in the rejection of  background events from $^{40}$K decay. 

A DU prototype, hosting three DOMs was operated for over a year at the KM3NeT-Italy site.  
From the data collected during this period~\cite{PPM-DU} a fast and reliable procedure for the time calibration which exploits $^{40}$K decays from the salt in the seawater and the LED
beacons inside the DOM has been tested. 

For the calibration of the PMT responses in the DOM, coincidences of Cherenkov photons from the $^{40}$K decay are used.  The PMTs time offsets, the relative efficiencies and the time spreads are extracted with this method with a fast code that can be easily expanded to the full detector. 
The time offsets between different DOMs are evaluated using the nanobeacons mounted inside the optical modules and cross checked with a calibration procedure using the signals from muons.  

With this time calibration method a precision at the level of nanosecond has been reached~\cite{PPM-DU}. 

\section{Status}

\begin{figure}
\includegraphics[clip=true,trim={0cm 0cm 0cm 0cm},width=0.5\textwidth]{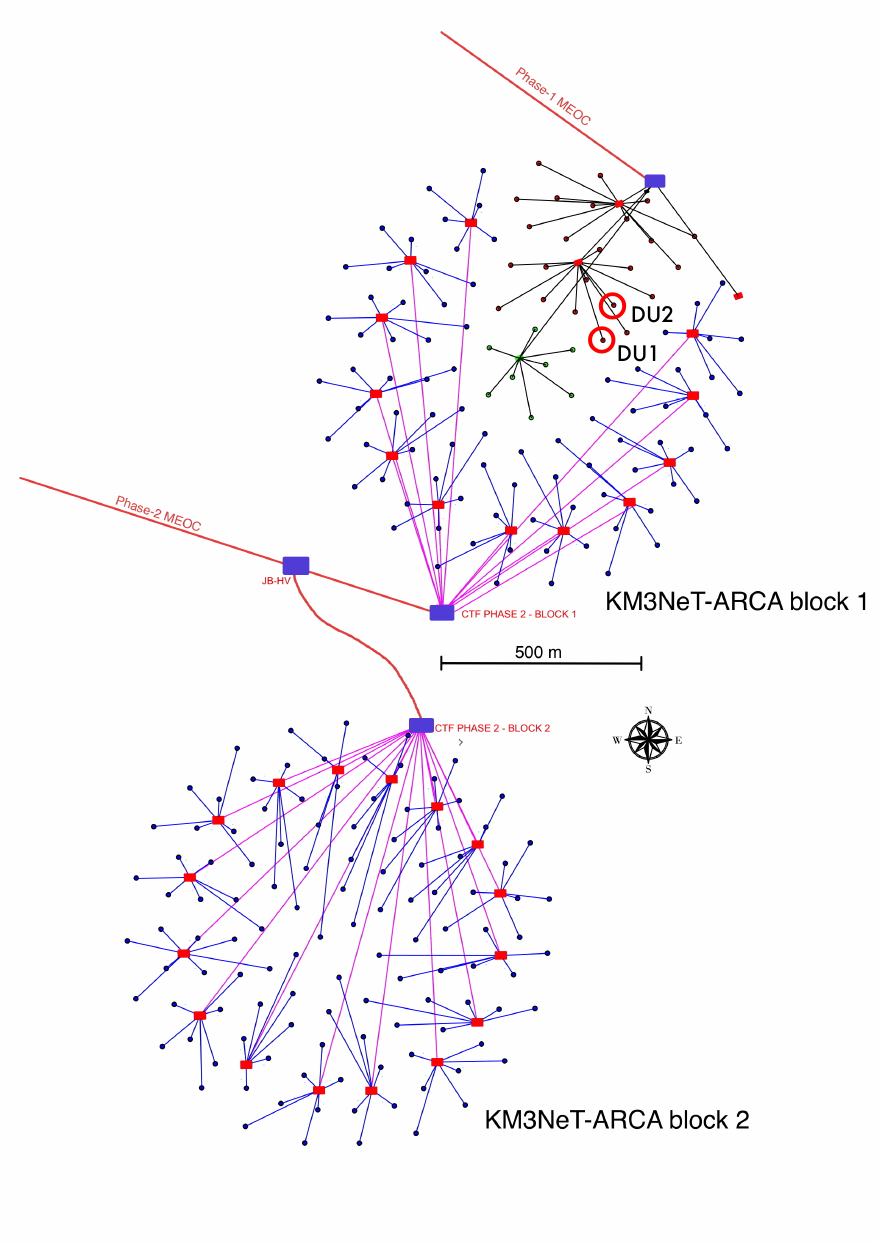}
\caption{Layout of the two ARCA building blocks. The schematic map of the cables connecting the DUs with the main cables is also reported. In black the positions and sea floor network of the ARCA phase-1. The two red circles indicate the positions of the two strings already deployed.}
\label{footprint}
\end{figure}

The ARCA detector, that will consist of two building blocks of 115 DUs, when finished will have the geometrical volume of  about 1 km$^{3}$. 
In Figure~\ref{footprint} the footprint of the detector is shown. 

Neutrino telescopes, being modular detectors, allow for staged construction and data analysis. The KM3NeT-phase1 will consist of 24 DUs string-like and 8 DUs based 
on a flexible tower design ~\cite{Tower_DIR,Tower_LongTerm} as first part of the ARCA detector (see Figure~\ref{footprint}) and 7 string-like DUs at the French site
with a denser distribution of the optical sensors, representing
the first nucleus of the ORCA detector. KM3NeT-phase1 has been funded and
the construction and deployments of the first strings started.

\begin{figure}
\includegraphics[clip=true,trim={0cm 0cm 0cm -2cm},width=0.5\textwidth]{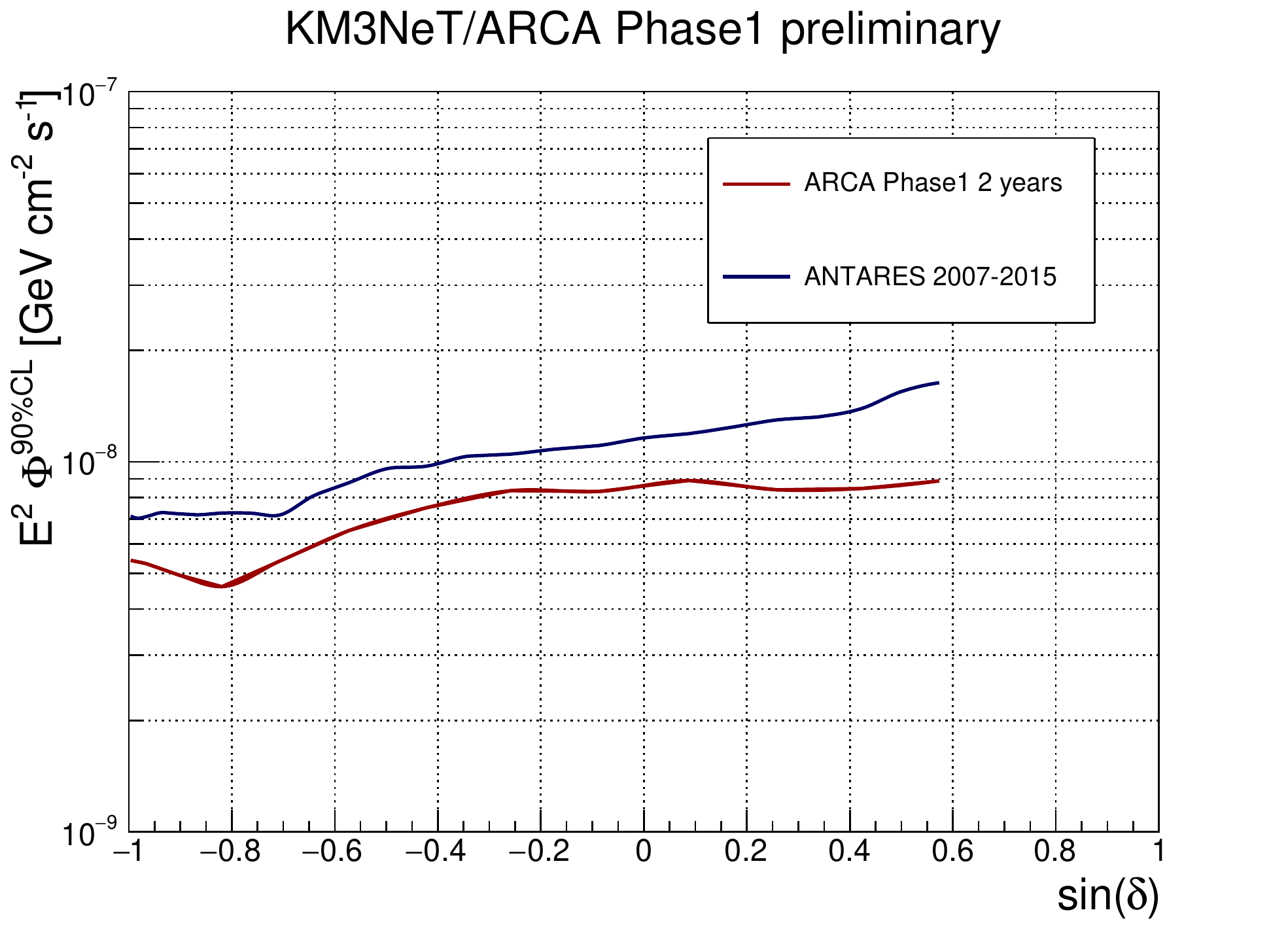}
\caption{ ARCA-phase1 (24 strings) sensitivity as a function of the source declination
(red line) for one neutrino flavour, for point-like sources with a spectrum E$^{-2}$ and 2 years of data-taking.
For comparison, the most recent (8 years of data taking) sensitivity of the ANTARES detector (blue line) is also shown.}
\label{Sensi}
\end{figure}

The ARCA phase1 detector (24 DUs string-like) when completed will be have a volume of about 0.1 km$^3$ and will quickly exceed the ANTARES sensitivity. In Figure~\ref{Sensi} the sensitivity for point-like source with spectrum proportional to E$^{-2}$ for two years of observation time is reported as a function of the source declination for ARCA phase1 (red continuos line). The sensitivity is evaluated for up-going track events. For comparison the best sensitivity for up-going track events of the ANTARES detector, reached after 8 years of observation time, is also reported (blue line). 

Currently two lines are in operation at Capo Passero site: the first one deployed in December 2015 and the second one in May 2016 (DU1 and DU2 respectively in Figure~\ref{footprint}). 
The analysis of the data taken with the two first lines started and currently the calibration procedures described in~\cite{PPM-DU} are on-going. 

\begin{figure}
\includegraphics[clip=true,trim={0cm 0cm 0cm 0cm},width=0.5\textwidth]{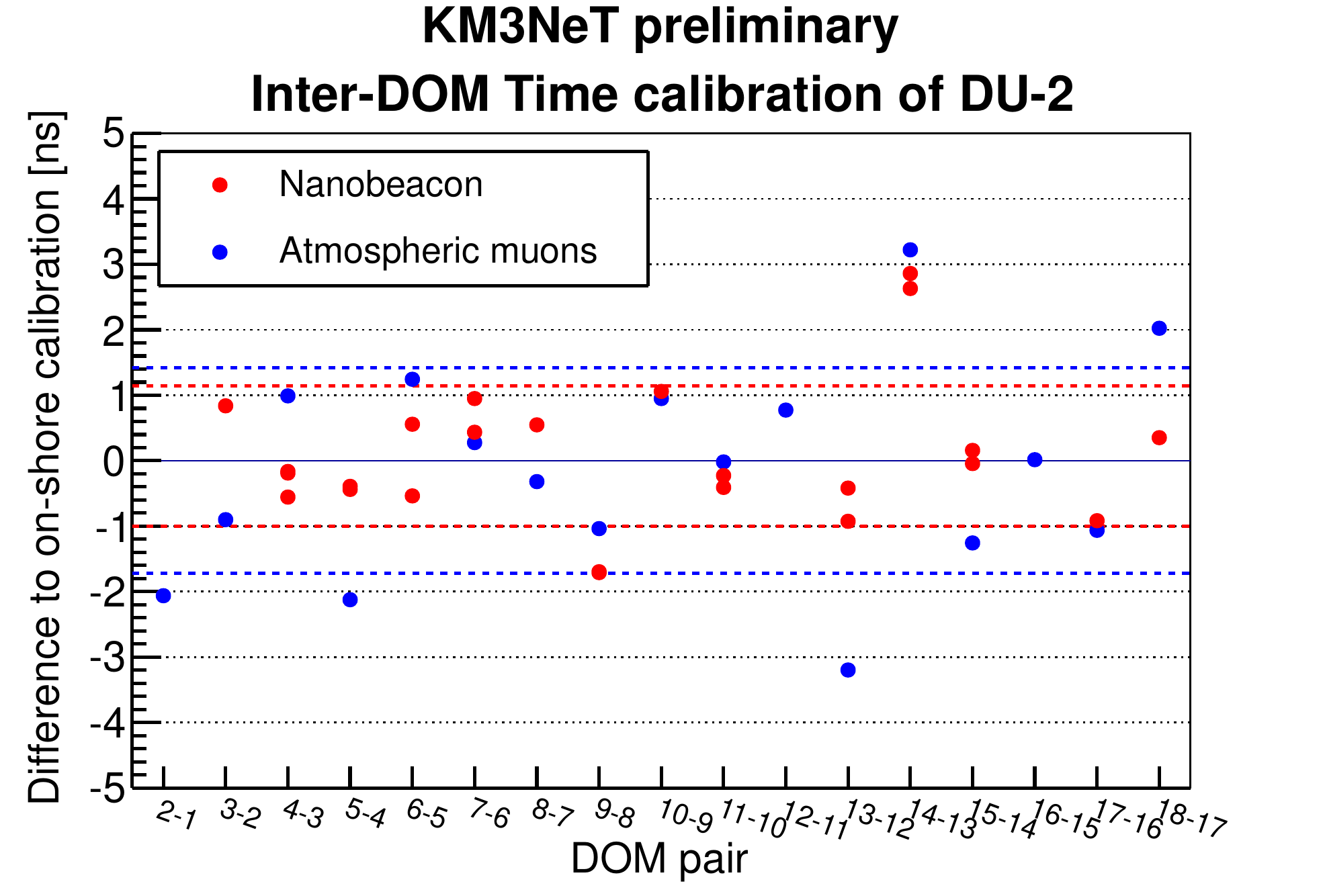}
\caption{Time differences between the time offsets estimated with the nanobeacons (red points) and muon signals (blue points) and the time offsets estimated on-shore. The dashed lines represent the standard deviation of the two distributions.}
\label{nanobeacon}
\end{figure}

Preliminary results of the inter-DOM time calibration performed with the nanobeacons are reported for the DU2 in Figure~\ref{nanobeacon} and compared with the calibration that uses the signals from muons. In particular the difference in time of the off-shore time offsets with the time offsets estimated on-shore   are reported. The on-shore time offsets have been estimated with a laser beam during the integration phase of the DU.
The results show that the two methods are in agreement and that the difference between the nanobeacons and the laser calibration is about 1.1~ns (standard deviation) while the calibration with atmospheric muon signals shows is around 1.5~ns.   

\begin{figure}
\includegraphics[width=0.5\textwidth]{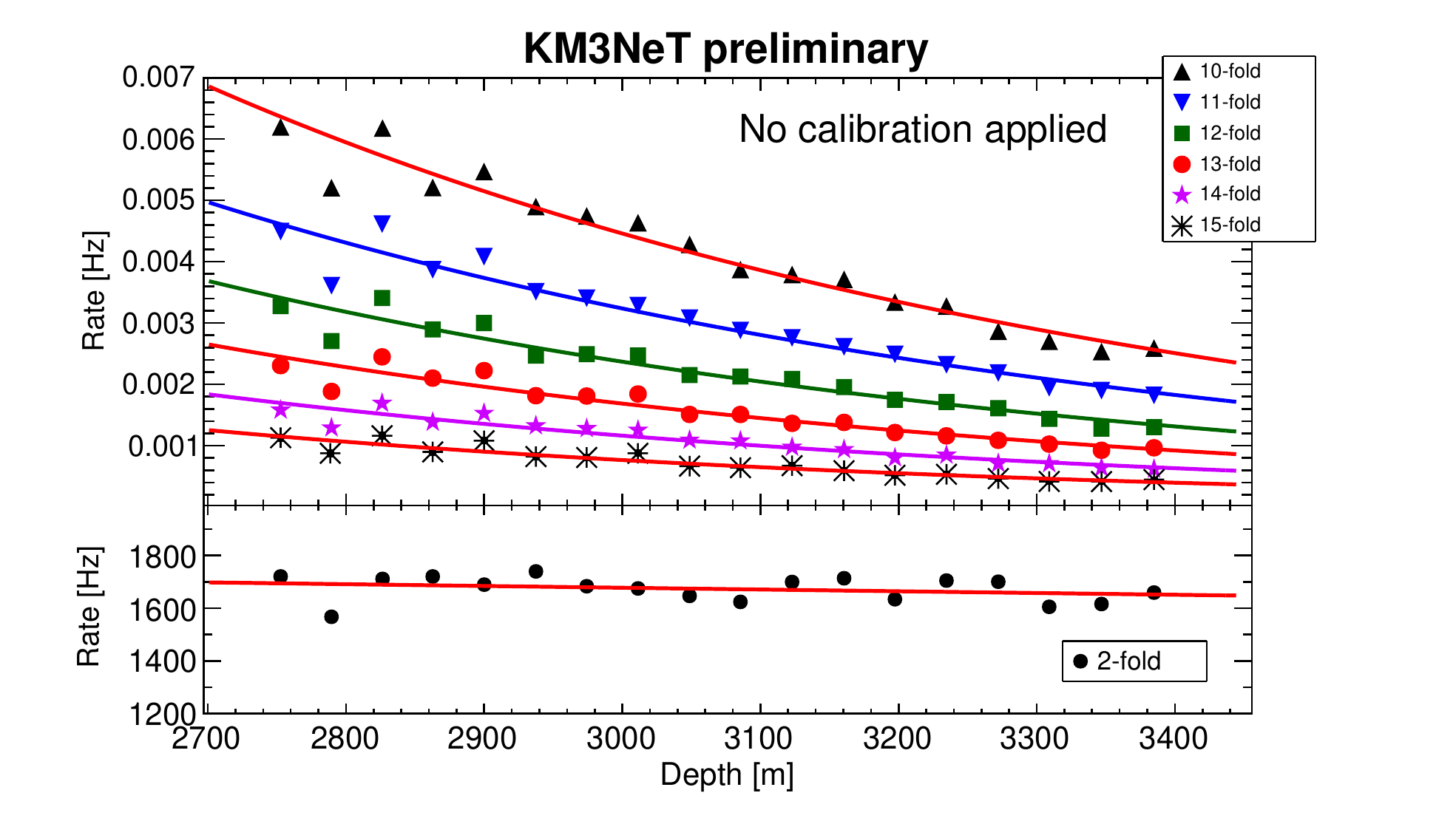}
\caption{Dependence of the event rates measured in the DOMs of DU1 as a function of its position with respect to the sea level (depth) for different coincidence levels (see text). Continuos lines represent the exponential fit. }
\label{Rate}
\end{figure}

With a very simple analysis of raw data the expected exponential dependence of the atmospheric muon rate as function of the depth has been found as shown in Figure~\ref{Rate}. This dependence has been obtained imposing cuts on the number of PMTs inside the single DOM with signal in a 20 ns time window  (coincidence level). In the bottom panel of Figure~\ref{Rate} this dependence is reported for low coincidence level (fold=2). The observed flat behaviour is consistent with the expected dependence of events due to $^{40}K$.  
%

%

\section{Summary}

In this contribution the main technical components
and the status of the KM3NeT/ARCA detector has been briefly presented. The construction of the detector, which
in its final configuration consists of 2 building blocks of 115 DUs with a total instrumented volume of
about 1 km$^3$, has begun. The detector sensitivities to neutrino fluxes from Galactic sources and to the diffuse flux measured
by IceCube have been discussed for the full ARCA detector.

KM3NeT-Phase1 is fully funded and will consist of 24 strings and 8 towers,
at the Italian site near Capo Passero (Sicily), and 7 strings at the French site, close to the Toulon coast. 
From MC simulation has been estimated that after two years of data taking ARCA phase1 will exceed the current ANTARES sensitivity to point-like sources (8 years of data taking).

The first two DUs of the ARCA detector have been deployed and currently the calibration phase is on going. 
Some preliminary results on raw data have been here presented showing the functionality of the detector.

\bigskip 

\end{document}